\documentclass{ws-ijmpc}

\usepackage[table,xcdraw]{xcolor}
\usepackage{graphicx,float}

\usepackage{tikz}
\usepackage{bm}
\usepackage{amsmath}
\usepackage{amssymb}
\usepackage{mathtools}
\usepackage[normalem]{ulem}

\usepackage{amsfonts}
\usepackage{combelow}
\usepackage{color}

\begin{document}

\markboth{G. Negro \& S. Busuioc \& V. E.  Ambru\c{s} 
\& G. Gonnella \& A. Lamura \& V. Sofonea}
{Comparison between isothermal collision-streaming and finite-difference lattice Boltzmann models}

\catchline{}{}{}{}{}
%%%%%%%%%%%%%%%%%%%%%%%%%%%%%%%%%%%%%%%%%%%%%%%%%%%%%%%%%%%%%%%%%%%%

\title{Comparison between isothermal collision-streaming and finite-difference lattice Boltzmann models}

\author{G. Negro}

\address{Dipartimento  di  Fisica,  Universit\`a  degli  Studi  di  Bari  and  INFN,\\
Sezione  di  Bari,  via  Amendola  173,  Bari,  I-70126,  Italy
}

\author{S. Busuioc}

\address{Department of Physics, West University of Timi\c{s}oara,\\ Bd. Vasile P\^{a}rvan 4, 
300223 Timi\c{s}oara, Romania}

\author{V. E.  Ambru\c{s}}

\address{Department of Physics, West University of Timi\c{s}oara,\\ Bd. Vasile P\^{a}rvan 4, 300223 Timi\c{s}oara, Romania}

\author{G. Gonnella}

\address{Dipartimento  di  Fisica,  Universit\`a  degli  Studi  di  Bari  and  INFN,\\
Sezione  di  Bari,  via  Amendola  173,  Bari,  I-70126,  Italy}

\author{A. Lamura}

\address{Istituto Applicazioni Calcolo, CNR,\\ Via Amendola 122/D, I-70126 Bari, Italy}

\author{V. Sofonea}

\address{Center for Fundamental and Advanced Technical Research, Romanian Academy,\\ Bd. Mihai Viteazul 24, 300223 Timi\c{s}oara, Romania}

\maketitle

\begin{history}
\received{Day Month Year}
\revised{Day Month Year}
\end{history}

\begin{abstract}

We present here a comparison between collision-streaming  
and finite-difference lattice Boltzmann (LB) models. 
This study provides a derivation of useful formulae which help one to properly
compare the simulation results obtained with both LB models. 
We consider three physical problems: the shock wave propagation, the damping of shear waves, and the decay of  Taylor-Green vortices, often used as benchmark tests. 
Despite the different mathematical and computational complexity 
of the two methods, we show how the physical results can be related to obtain relevant quantities. 

\end{abstract}

\ccode{PACS Nos.: }

\section{Introduction}
Since more than 3 decades ago, the use of lattice Boltzmann (LB) 
models to address hydrodynamic problems has widely expanded because of the
parallel nature of their basic algorithm, as well as of their 
capability to easily handle interparticle interactions and boundary conditions\cite{fluid.30.1.329,krueger17,succi18}. 
A characteristic feature of the LB models is the polynomial expansion\cite{shanJFM2006}
of the single-particle equilibrium  distribution function
$f^{eq}(\bm{x}, \bm{v}, t)$ up to a
certain order $N$ with respect to the fluid velocity  $\bm{u} \equiv\bm{u}(\bm{x},t)$.
A rigorous way to do this, is projecting the equilibrium distribution
function on a set of orthogonal polynomials. In addition, the use of Hermite 
polynomials as the expansion basis has the unique feature that the 
expansion coefficients correspond precisely to the velocity
moments\cite{shanJFM2006} up to a given degree. 
The polynomial expansion and the application of the Gauss quadrature theorem allows one to
compute the moments of $f^{eq}(\bm{x}, \bm{v}, t)$,
as well as of the distribution function $f(\bm{x}, \bm{v}, t)$,
which appear in the Boltzmann equation, 
by summation over a discrete velocity set $\bm{v}_{k}$,
$1 \leq k \leq K$.\cite{fluid.30.1.329,krueger17,succi18,shanJFM2006}

In the widely used \emph{collision-streaming} (CS) lattice Boltzmann
models, the velocity space is discretized so that the velocity
vectors of the fluid particles leaving a node of the lattice are
oriented towards the neighboring
nodes.\cite{fluid.30.1.329,krueger17,succi18} Such models are
also called on-lattice models. 
%%%%%%%%%%%%%%%%%%%%%%%%%%%%%%%%%%%%%%%%%%%%%%%%%%%%%%%%%%%%%
Alternatively, in the \emph{finite-difference} (FD) lattice Boltzmann 
models\cite{biciuscaCRM2015,sofoneaPRE2018}, the velocity vectors ${\bm{v}}_k$  are 
 obtained using the Gauss-Hermite quadrature method in the velocity space.
These vectors are generally off-lattice and their Cartesian
components are
expressed as irrational numbers, namely the roots of the Hermite polynomials. 
For this reason, in these models the distribution functions are evolved using 
an appropriate finite-difference scheme. 
\cite{sofoneaPRE2018,CRISTEA2006113,articlecristeagravity}

To the best of our knowledge, no comparison is available
in the literature between the two schemes to properly match
physical quantities in terms of the model parameters.
In this study, we aim to provide the tools necessary to perform 
such comparisons, which may be used to compare, for example, past 
results obtained for nonideal fluids using various LB models.
Indeed, in the past both CS \cite{doi:10.1063/1.4937595,PhysRevE.89.063303} and 
FD \cite{sofoneaPRE2018,CRISTEA2006113,sofoneaPRE2004,gonnellaPRE2007} 
lattice Boltzmann models were used to study liquid-vapor systems. 
A comparison between the two methods lacked and no attempt was done so far to have
a unified framework to map models onto each other. 

The outline of this paper is as follows.
In Sec.~\ref{sec:LB}, we briefly describe the CS and the FD 
lattice Boltzmann models here considered.
In Sec.~\ref{sec:adim} we derive the way enabling the
conversion between the non-dimensionalization procedures
currently used in these models. 
This ensures the simulation of the same physical system
with the two models.
In Sec.~\ref{sec:res}, we compare the two LB models (CS and FD) by
considering simple problems involving an ideal fluid
under the assumption of isothermal 
conditions. Our conclusions are summarised in Sec.~\ref{sec:conc}.

\section{Lattice Boltzmann models}\label{sec:LB}
%%%%%%%%%%%%%%%%%%%%%%%%%%%%%%%%%%%%%%%%%%
%%%% TIMISOARA %%%%%%%%%%%%%%%%%%%%%%%%%%%
%%%%%%%%%%%%%%%%%%%%%%%%%%%%%%%%%%%%%%%%%%

When the Bhatnagar-Gross-Krook (BGK) collision term is used
in an isothermal LB model, the moments of the distribution function
$f({\bm{x}},{\bm{v}},t)$ up to order $N=2$ are needed in order to get 
the evolution equations of the macroscopic fields at
the incompressible Navier - Stokes level
\cite{fluid.30.1.329,krueger17,succi18,shanJFM2006,GuoShuLBapplications,ambrusPRE2012}. 
The minimum number of the velocity vectors in the
two-dimensional ($D=2$) isothermal LB model based on the
full-range Gauss-Hermite quadrature ensuring all the moments of
$f({\bm{x}},{\bm{v}},t)$ up to order $N=2$ is $K=(N+1)^D=9$.\cite{shanJFM2006,fedeIJMPF2015,ambrus16jcp,piaudIJMPC2014}

As usual in the current LB models involving the BGK collision
term,\cite{shanJFM2006} the non-dimensionalized form of
the evolution equation of the functions  $f_k \equiv f(\bm{x}, \bm{v}_k, t)$ for the force-free flow of a
single-component fluid is 
\begin{equation}\label{evolution}
 \partial_{t} f_k +
 {\bm{v}}_k\cdot \nabla f_k 
% + \bm{F} \cdot  (\nabla_{\bm{p}} f)_{\bm{\kappa}}
 = -\frac{1}{\tau}[f_k - f^{eq}_k],
\end{equation}
where $\tau$ is the non-dimensionalized value of the 
relaxation time. For simplicity, in this paper the value of $\tau$ 
is assumed to be constant.
The details of the non-dimensionalization 
procedure used in the FD and the CS lattice Boltzmann models are 
discussed in Section 3 below.
The equilibrium single-particle distribution function, expanded up to 
order $N=2$, is given by:
\begin{equation}
f^{eq}_{k}  =  w_{k}\rho \left\{\,1\,+\,{\bm{v}}_{k}\cdot {\bm{u}} +
\frac{1}{\,2\,} \left[ ({\bm{v}}_{k}\cdot {\bm{u}})^{2} - u^{2} \right] \right\}
\label{herfeq}
\end{equation}
After the aforementioned discretization of the velocity space, the macroscopic quantities,
namely the fluid density $\rho$ and momentum density $\rho\bm{u}$, are computed as
\begin{equation}
 \begin{pmatrix}
  \rho \\ \rho\bm{ u}
 \end{pmatrix} = 
 \sum_k f_k 
 \begin{pmatrix}
  1 \\ \bm{v}_k
 \end{pmatrix}.
 \label{eq:f_moments_discrete}
\end{equation}

When using the finite-difference LB model in this paper, the evolution equation (\ref{evolution}) is solved by
using the third order total variation diminishing (TVD) Runge-Kutta (RK-3) time stepping procedure,\cite{gottlieb98,henrick05,shu88,trangenstein07}
together with the fifth-order weighted essentially non-oscillatory (WENO-5) scheme for the
advection.\cite{gan11,jiang96,camwa2019}

Using the Chapman-Enskog method, it can be shown 
that, when the fluid satisfies the Navier-Stokes equation, the
non-dimensionalized value of the kinematic viscosity is given by
\begin{equation}\label{eq:nu_fd}
\nu_{\text{FD}}=\frac{\tau_{\text{FD}} T}{m}
\end{equation}
where $\tau_{\text{FD}}$ is the relaxation time non-dimensionalized with 
respect to the finite difference conventions discussed in Sec.~\ref{sec:adim},
$T$ is the non-dimensionalized value of the 
local fluid temperature and $m$ is the non-dimensionalized value of the
fluid particle mass.

%%%%%%%%%%%%%%%%%%%%%%%%%%%%%%%%%%%%%%%%%%%%%%%%%%
%%%%%%%%%%%% BARI %%%%%%%%%%%%%%%%%%%%%%%%%%%%%%%%
%%%%%%%%%%%%%%%%%%%%%%%%%%%%%%%%%%%%%%%%%%%%%%%%%% 

In the collision-streaming LB models, the
fluid particles collide in the lattice 
nodes and thereafter move in the time lapse $\delta t$ towards the
neighboring nodes, with speed $c_l=\delta s / \delta t$
along the lattice links of spacing $\delta s$. 
The distribution functions follow the governing
equation (in the BGK approximation):
\begin{equation}
f_k({\bm{x}}+{\bm{v}}_k\delta t,t+\delta t) -
f_k({\bm{x}},t)=-\frac{\delta t }{\tau}\left[f_k({\bm{x}},t)-f_k^{\textrm{eq}}
({\bm{x}},t)\right] \label{sc_evolution} \ ,
\end{equation}
where $\{\bm{v}_k\}$, $0\le k\le K-1$, is the set of discrete 
velocities. The equilibrium functions $f_k^{\textrm{eq}}({\bm{x}},t)$ are given by a second order
expansion of the Maxwell-Boltzmann distribution function with
 respect to the Hermite
polynomials.\cite{doi:10.1063/1.4937595,PhysRevE.89.063303}
In all the CS simulations, the non-dimensionalised values $(\delta s)_{\text{LU}}=1$ and $(\delta t)_{\text{LU}} = \sqrt{3}/3$ were used to fix the non-vanishing Cartesian projections $v_{k;\alpha}$ ($\alpha = 1,\, \ldots D$,
$\vert v_{k;\alpha} \vert = c_{l}$)   of the vectors $\bm{v}_{k}$, such that $c_{l} = \sqrt{3}$, as prescribed by the Gauss -Hermite quadrature on the $D2Q9$ 
lattice.\cite{PhysRevE.89.063303}

In the CS lattice Boltzmann model, the relaxation time $\tau$ controls the kinematic viscosity
\begin{equation}
\nu_{\text{LU}}=\left(\tau_{\text{LU}}-\frac{\delta t}{2}\right),
\label{nulu}
\end{equation}
where the subscript LU stands for lattice units. 

\section{Relations between non-dimensionalization conventions}\label{sec:adim}

In order to relate the non-dimensional values for a quantity 
$\tilde A$ (the tilde indicates a dimensional quantity), obtained 
using two non-dimensionalization conventions ($A_1$ and $A_2$), 
the following formula can be used\cite{YEOMANS2006159}:
\begin{equation}
A_1=A_2\frac{\tilde A_{\textrm{ref};2}}{\tilde A_{\textrm{ref};1}}\ ,
\end{equation}
since $\tilde{A} = A_1 \  \tilde A_{\textrm{ref};1}= A_2 \ \tilde A_{\textrm{ref};2}$. 

We wish to simulate the same fluid system using both the FD and the CS lattice Boltzmann models. Since
the reference values used in these models may be different, but the
computer simulations are usually performed using non-dimensionalized 
quantities, we need the conversion relations between the
non-dimensionalized values of the physical quantities used to describe
the fluid properties and the flow geometry within each model.
In the sequel, we will use the subscripts FD and LU to denote the 
physical quantities in the FD and the 
CS models, respectively. We choose to use LU (which stands for 
"lattice units") since this notation is frequently encountered
in the LB literature dealing with CS models.

Let us consider a fluid system whose characteristic length is $\widetilde{L}$, in which an ideal fluid 
with viscosity $\widetilde{\nu}$ is maintained at the constant temperature
$\widetilde{T}_0 = T_0 \widetilde{T}_{\textrm{ref}}$. 
In this paper, we assume that the reference temperature
$\widetilde{T}_{\textrm{ref}}$, the reference pressure
$\tilde{P}_{\textrm{ref}}$,
the reference mass $\widetilde{m}_{\rm ref}$,
as well as
the reference density $\tilde{\rho}_{\textrm{ref}}$ are
identical in both the
CS and the FD models.

The reference speed in the two models is:
\begin{equation}
 \widetilde{c}_{\rm ref; LU} =  \widetilde{c}_{\rm ref; FD} = 
 \sqrt{\frac{ \widetilde{K}_B \widetilde{T}_{\rm ref}}{\widetilde{m}_{\rm ref}}},
 \label{eq:cref}
\end{equation}
where $\widetilde{K}_{B}$ is the Boltzmann constant, and $\widetilde{T}_{\rm ref}$ is the reference temperature in both models.

Let the reference length in the FD approach be the system size 
$\widetilde{L}_{\rm ref; FD} = \widetilde{L}$, while in 
the CS approach, it is the lattice spacing. 
Considering that the CS simulation is performed on a lattice 
containing $N_{\text{LU}}$ nodes along the characteristic length
$\tilde{L}$, the reference length in the CS model is
\begin{equation}
 \widetilde{L}_{\rm ref; LU} = \frac{\widetilde{L}}{N_{\rm LU}} = 
 \frac{\widetilde{L}_{\rm ref; FD}}{N_{\rm LU}}.
\end{equation}
The reference time in the FD approach is
\begin{equation}
 \widetilde{t}_{\rm ref; FD} = \frac{\widetilde{L}}{\widetilde{c}_{\rm ref; FD}}.
\end{equation}
The reference time in the LU approach is:
\begin{equation}
 \widetilde{t}_{\rm ref; LU} = \frac{\widetilde{L}}{N_{\rm LU} \widetilde{c}_{\rm ref; LU}} 
 = \frac{\widetilde{t}_{\rm ref; FD}}{N_{\rm LU}}.
\end{equation}

In order to ensure that the same system is being simulated, the viscosity must be fixed.
The reference viscosity in the FD approach is:
\begin{equation}
 \widetilde{\nu}_{\rm ref; FD} = \frac{\widetilde{t}_{\rm ref; FD}\widetilde{P}_{\rm ref}}{\widetilde{\rho}_{\textrm{ref}}}  
 = \frac{\widetilde{L} \widetilde{P}_{\rm ref}}{\widetilde{c}_{\rm ref; FD}\widetilde{\rho}_{\textrm{ref}}},
\end{equation}
being independent of the simulation details, such as number of nodes or time step,
where $\widetilde{P}_{\rm ref}$ is the reference pressure.
The LU reference viscosity reads:
\begin{equation}
 \widetilde{\nu}_{\rm ref; LU} = \frac{\widetilde{\nu}_{\rm ref; FD}}{N_{\rm LU}}.
\end{equation}
Thus, the LU reference viscosity depends on the number of lattice nodes $N_{\rm LU}$. 

This result, as well as the
expression (\ref{nulu}) of the non-dimensionalized viscosity
value in the CS model, allows us to get the
relation between the  non-dimensionalized
FD relaxation time $\tau_{FD}$ and the corresponding
value of $\tau_{LU}$:
\begin{equation}\label{eq:nu-transfer}
 \tau_{\rm LU} = \nu_{\rm LU} + \frac{\delta t_{\rm LU}}{2}
 = N_{\rm LU} \tau_{\rm FD} + \frac{\sqrt{3}}{6},
\end{equation}
where the last term represents the numerical correction typical for collision-streaming 
simulations.

\section{Numerical results}\label{sec:res}
We present here different standard physical-benchmark problems in order to compare the two models. 
In all simulations periodic boundary conditions were considered.

\subsection{Shock Waves}\label{sec:res:shock}
As a first test problem we consider the Cartesian shock problem.

The test consists of a one-dimensional Riemann problem:
In an isothermal ideal gas at temperature $T$, the
density is initialized as follows:
\begin{align}
\begin{cases}
\rho(x)&=\rho_{\rm L} \ \ \ \textrm{if} \ \ \  x\le x_0\\
\rho(x)&=\rho_{\rm R} \ \ \ \textrm{otherwise} \ ,
\end{cases}
\end{align}
where $\rho_{\rm L}$ and $\rho_{\rm R}$ are the values of the 
density to the left and to the right of the initial discontinuity,
which is located at $x = x_0$.
Since in our simulation setup, the density is related to the pressure $P$
through $\rho = m P / T$, where $T$ is considered to be constant, 
we expect no contact discontinuity to appear in our simulation results.
This can be seen by considering the Euler equations,
reproduced below for the one-dimensional flow of an isothermal fluid:
\begin{equation}
 \partial_t \rho + \partial_x (\rho u) = 0, \qquad
 \partial_t (\rho u) + \partial_x(\rho u^2 + P) = 0.
 \label{eq:euler}
\end{equation}
Introducing the similarity variable 
\begin{equation}
 \xi = \frac{x - x_0}{t},
\end{equation}
it can be seen that Eq.~\eqref{eq:euler} reduces to:
\begin{equation}
 \frac{\partial u}{\partial \xi} = \frac{\xi - u}{\rho} 
 \frac{\partial \rho}{\partial \xi}, \qquad 
 \frac{\partial P}{\partial \xi} = 
 (\xi - u)^2 \frac{\partial \rho}{\partial \xi}.
 \label{eq:euler_xi}
\end{equation}
Noting that $P = \rho c_s^2$, 
where $c_s = \sqrt{T / m}$
is the non-dimensionalised speed of 
sound in an isothermal fluid, the above equations are 
satisfied either when $\rho$ and $u$ are constant, or when 
\begin{equation}
 u = \xi \pm c_s.
\end{equation}
The above solution corresponds to a rarefaction wave 
travelling to the left ($+$) or to the right ($-$).
We note that the solution $u = \xi$ (corresponding to the 
contact discontinuity) does not appear in the case of isothermal 
flows.

Assuming that $\rho_{\rm L} > \rho_{\rm R}$, the rarefaction wave 
propagates to the left, in which case the velocity can be seen to 
increase linearly according to:
\begin{equation}
 u_*(\xi_*) = \xi_* + c_s,\label{eq:shock_ustar}
\end{equation}
where the star ($*$) is employed to indicate that the 
analysis is restricted to the rarefaction wave. 
From Eq.~\eqref{eq:shock_ustar} it can be seen that the 
head of the rarefaction wave travels with constant 
velocity 
\begin{equation}
 \xi_r = -c_s.
\end{equation}
The tail of the rarefaction wave corresponds to the 
value $\xi_c$ of the similarity variable, for which 
the velocity takes the constant value on 
the plateau, $u = u_c$:
\begin{equation}
 \xi_c = c_s (\zeta - 1), \qquad 
 \zeta = \frac{u_c}{c_s},\label{eq:shock_xic}
\end{equation}
where the dimensionless quantity $\zeta$ was introduced for 
future convenience.
The value of $u_c$ will be determined further below.

Inserting Eq.~\eqref{eq:shock_ustar} into Eq.~\eqref{eq:euler_xi} 
gives the solution
\begin{equation}
 \rho_*(\xi_*) = \rho_{\rm L} \exp\left(-\frac{\xi_* - \xi_r}{\xi_r}\right) 
 = \rho_{\rm L} \exp\left[-\zeta \frac{u_*(\xi_*)}{u_c}\right].
 \label{eq:shock_rhostar}
\end{equation}
It can be seen that the density on the central plateau,
$\rho_c$, can be determined once $\zeta$ is known using the 
equation
\begin{equation}
 \rho_c = \rho_{\rm L} e^{-\zeta}.\label{eq:shock_rhoc}
\end{equation}

Let us now consider the Rankine-Hugoniot junction conditions
for a discontinuity having the similarity variable 
$\xi_s$:
\begin{equation}
 \rho_+ (u_+ - \xi_s) = \rho_- (u_- - \xi_s), \qquad 
 \rho_+ u_+ (u_+ - \xi_s) + P_+ = \rho_- u_- (u_- - \xi_s) + P_-,
 \label{eq:shock_rh}
\end{equation}
where $+$ and $-$ denote the fluid properties to the 
right and to the left of the discontinuity, respectively. 
Specializing the above equations to the case of the 
shock front, where $\rho_+ = \rho_{\rm R}$ and $u_+ = 0$,
the following relations are obtained:
\begin{equation}
 \xi_s = \frac{\rho_c \zeta c_s}{\rho_c -\rho_{\rm R}}, \qquad 
 \rho_{c} - \zeta^2
 \frac{\rho_c \rho_{\rm R}}{\rho_c - \rho_{\rm R}} - \rho_{\rm R} = 0.
 \label{eq:shock_rh_aux}
\end{equation}
Inserting $\rho_c$ from \eqref{eq:shock_rhoc} in the above relations, 
the value of $\zeta$ can be found by solving the following 
nonlinear equation:
\begin{equation}
 2 + \zeta^2 - \frac{\rho_{\rm L}}{\rho_{\rm R}} e^{-\zeta} - 
 \frac{\rho_{\rm R}}{\rho_{\rm L}} e^{\zeta} = 0 .
\end{equation}
In order to obtain the full solution,
the value of $\zeta$ must be inserted in Eqs.~\eqref{eq:shock_xic} and 
\eqref{eq:shock_rhoc} to obtain the velocity $\xi_c$ of the tail of the 
rarefaction wave and the density $\rho_c$ of the central plateau. 
The velocity $\xi_s$ of the shock front can be obtained from 
Eq.~\eqref{eq:shock_rh_aux}:
\begin{equation}
 \xi_s = \frac{\zeta c_s}{1 - \frac{\rho_{\rm R}}{\rho_{\rm L}} e^\zeta} .
\end{equation}

\begin{figure}[t]
\begin{tabular}{c}
\includegraphics[width=0.85\linewidth]{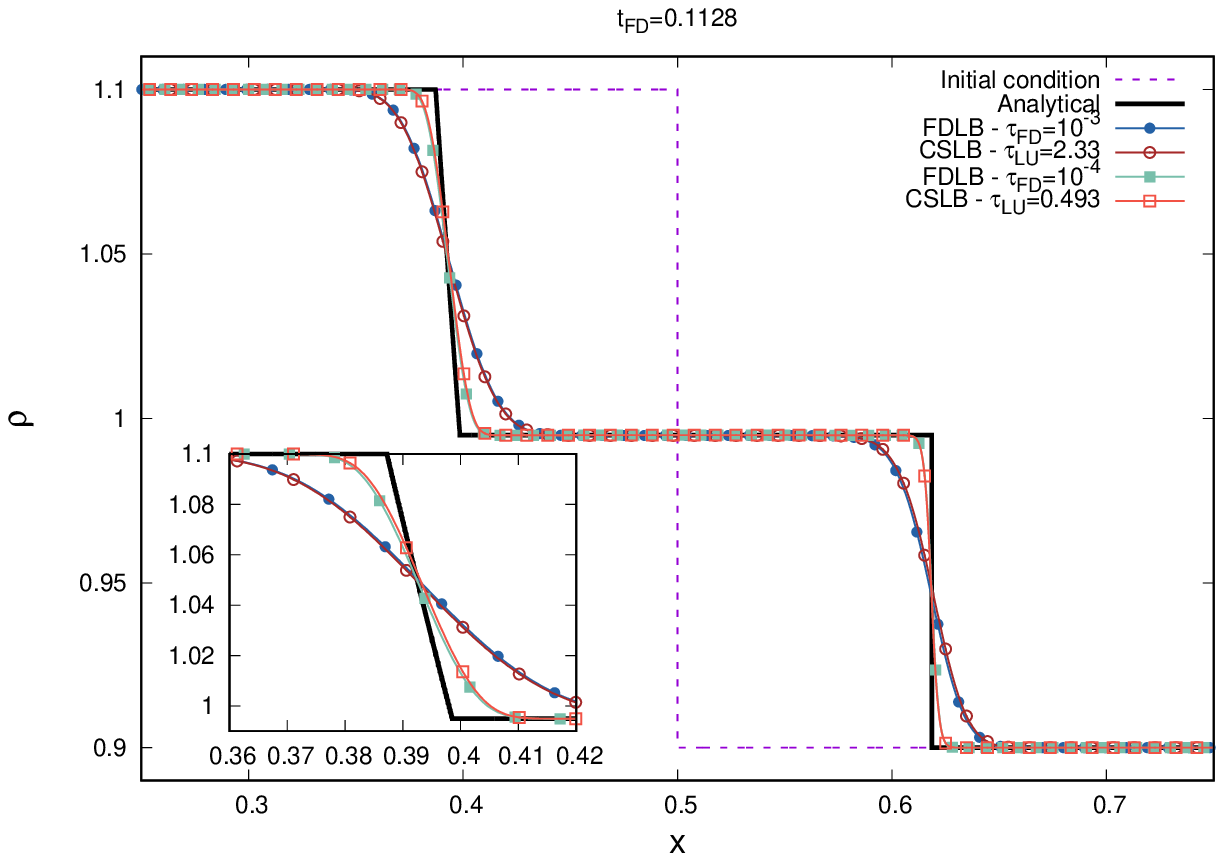}\\
\includegraphics[width=0.85\linewidth]{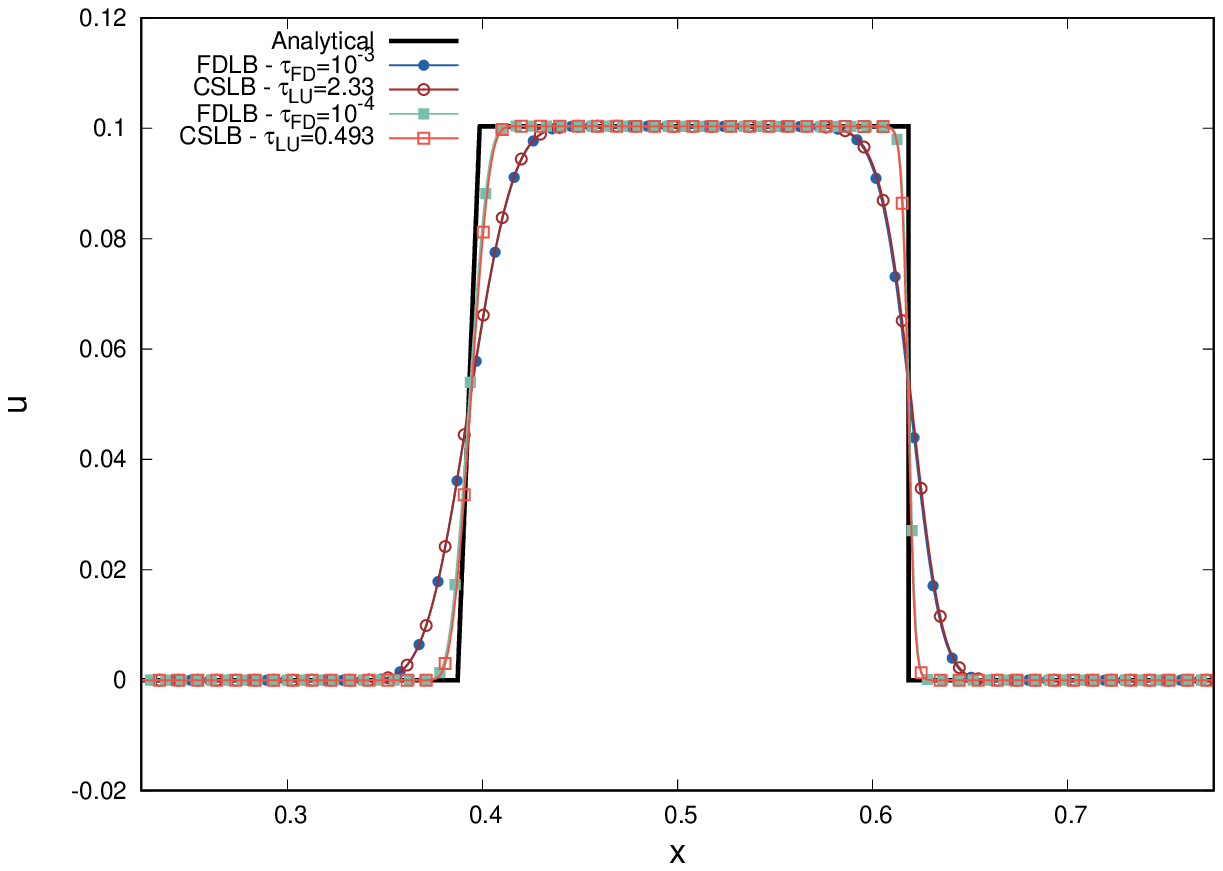}
\end{tabular}
\caption{Comparison of FDLB and CSLB results in the context of the Cartesian shock problem
at the level of the density \textbf{(upper panel)} and velocity \textbf{(lower panel)} profiles, 
obtained at $t_{\text{FD}} \simeq 0.1128$ for various values of the relaxation time.
The \textbf{inset} shows the shock front.}
\label{f1}
\end{figure}

We now discuss our numerical results. 
We consider that the fluid temperature is the reference 
temperature, such that $T=T_{\text{LU}}=T_{\text{FD}}=1$. 
Hence the non-dimensionalized sound speed 
in both LB models is $c_{s} = 1$.
In order to reduce the errors due to compressibility effects, we 
take $\rho_{\rm L} = 1.1$ and $\rho_{\rm R} = 0.9$, where 
the reference density is taken to be the average of 
$\rho_{\rm L}$ and $\rho_{\rm R}$.
In this case, $\zeta \simeq 0.10035$ and the 
relevant finite difference quantities are given below:
\begin{equation}
 \xi_r = -1, \quad \xi_c \simeq -0.900, \quad u_c = \zeta \simeq 0.10035, 
 \quad \rho_c \simeq 0.995, \quad \xi_s \simeq 1.051.
\end{equation}

The discontinuity in density makes the simulation of 
shock waves propagation a good test for the numerical methods used.
The initial density jump creates a density wave traveling from high density regions to lower density ones.  
We fixed the number of nodes at $N_x=N_{LU}=2048$ and considered two values of the relaxation time, namely $\tau_{\text{FD}}=\{10^{-4},10^{-3}\}$, 
corresponding to $\tau_{\text{LU}}=\{0.493,2.33\}$. In Fig.~\ref{f1}, the density and velocity profiles obtained with the two methods are represented 
at time $t_{\text{FD}} \simeq 0.1128$ (attained after $400$ iteration using CSLB),
alongside the analytic solution for the inviscid case. 
The curves show good agreement between the two models for the considered values of viscosity.

In the CSLB implementation,
$(\delta t)_{\text{LU}} = 1/\sqrt{3}$ corresponds to the
time step $(\delta t)_{\text{FD}} = 1/N\sqrt{3} \simeq 2.82 \times 10^{-4}$.
The Courant-Friedrichs-Lewy number, ${\rm CFL} = c_l \delta t /\delta s$, is equal 
to one for this choice of parameters. 
In the FDLB implementation, the time step is bounded by the CFL condition
${\rm CFL} \le 1$, such that the maximum time step permitted is that employed 
in the CSLB implementation. The time step in the FDLB implementation 
is further restricted to obey $(\delta t)_{\text{FD}} < \tau_{\text{FD}}$, in order to prevent the 
collision term from becoming stiff. Thus, at 
$\tau_{\text{FD}} = 10^{-3}$, we performed the FDLB simulations using 
$(\delta t)_{\text{FD}} = 1/N\sqrt{3}$, while 
at $\tau_{\text{FD}} = 10^{-4}$, the time step was decreased by a factor of $3$,
$(\delta t)_{\text{FD}} = 1/3N\sqrt{3} \simeq 9.40 \times 10^{-5}$, such that 
$1200$ iterations were required to reach the state shown in Fig.~\ref{f1}. 
We note that the restriction $(\delta t)_{\text{FD}} < \tau_{\text{FD}}$ can be lifted, 
e.g., when implicit-explicit (IMEX) schemes are employed.\cite{wang07}

\subsection{Shear waves}\label{sec:res:shear}

In order to compare numerical viscosity effects in the two models,
we analyze in this subsection the evolution of
shear waves. We consider waves of wavelength $\lambda = 1$  in an ideal gas with density
$\rho=1$ at temperature $T=1$.

In the simulations performed, the wave vector
${\bm{k}}$, $\vert {\bm{k}} \vert = 2\pi/\lambda=2\pi$, was aligned 
along the horizontal axis and its Cartesian components were $(2\pi, 0)$.

\begin{table}[ht]
\tbl{Apparent kinematic viscosity $\nu_{app}$, expressed using the FD adimensionalization, measured as a numerical fit of Eq. (\ref{eq:shear_sol})
in the context of the damping of shear waves. }
%\centering
{\begin{tabular}{@{}|ll|ll|ll|@{}}
\hline
\multicolumn{2}{|l|}{\cellcolor[HTML]{656565}\ }   &
\multicolumn{2}{c|}{\cellcolor[HTML]{656565}{\color[HTML]{FFFFFF}CSLB}}   &
\multicolumn{2}{c|}{\cellcolor[HTML]{656565}{\color[HTML]{FFFFFF}FDLB}}   \\\hline
$\tau_{\text{LU}}$ & $N_x$ & $\nu_{app}$   & Rel. err. & $\nu_{app}$ & Rel. err. \\ 
$0.2986$    & $20$  & $0.0005073$ & $0.0154$  & $0.0005271$ & $0.0542$      \\
$0.3036$    & $30$  & $0.0005032$ & $0.0065$  & $0.0005039$ & $0.0078$      \\
$0.3086$    & $40$  & $0.0005017$ & $0.0034$  & $0.0005009$ & $0.0018$      \\
$0.3136$    & $50$  & $0.0005015$ & $0.0021$  & $0.0005003$ & $0.0006$     \\
$0.3186$    & $60$  & $0.0005006$ & $0.0013$  & $0.0005001$ & $0.0002$      \\ 
\hline
\end{tabular}
\label{tabel_shear}}
\end{table}

Let ${\bm{u}}({\bm{x}},t)$ be the fluid velocity vector. 
In both series of simulations, the velocity field was
initialized according to:
\begin{subequations}
\begin{eqnarray}
u_{x}({\bm{x}},0) & = & 0, \\
u_{y}({\bm{x}},0) & = & U \sin({\bm{k}} \cdot {\bm{x}}),
\end{eqnarray}
\end{subequations}
with $U = 0.01$. When the fluid is not too far from the equilibrium (i.e.,
when the relaxation time is small enough), the fluid evolves according
to the Navier-Stokes equations. In the setup of the shear waves problem, we have
$u_{x}({\bm{x}},t) = 0$ and there is no
spatial variation of the velocity vector along the $y$ direction.
Under these circumstances and assuming that the fluid is isothermal and incompressible, 
the Navier-Stokes equations reduce to:
\begin{equation}
\partial_{t} u_{y}({\bm{x}},t)\,-\,\nu_{0}\,
\partial_{x}^{2}u_{y}({\bm{x}},t) = 0.
\end{equation}

Assuming that for $t > 0$, $u_y({\bm{x}},t) = \widetilde{u}(t) \sin ({\bm{k}} \cdot {\bm{x}})$, the solution is:
\begin{equation}
 \widetilde{u}(t) = U e^{-k^2 \nu_0 t},
 \label{eq:shear_sol}
\end{equation}
where $\nu_0$ is the analytic kinematic viscosity.

\begin{figure}
\begin{center}
\includegraphics[width=0.7\linewidth]{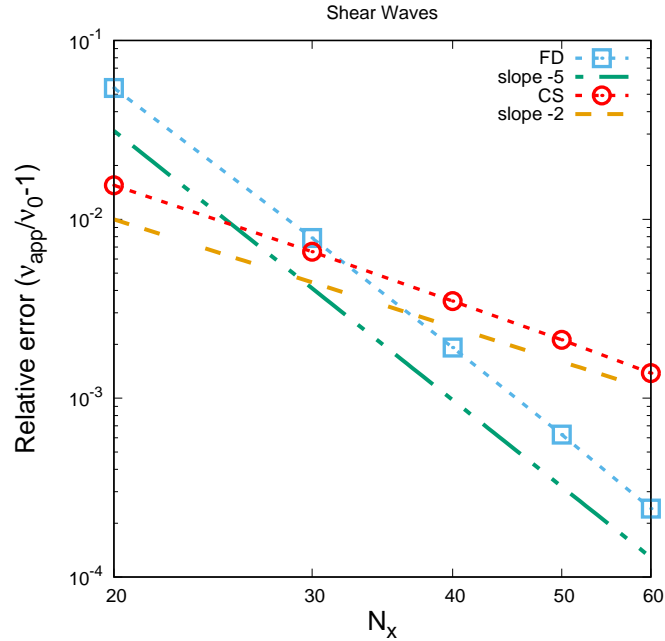}
\caption{Relative error of the measured kinematic viscosity $\nu_{app}$ expressed 
with respect to the expected analytic value $\nu_0$, extracted from the numerical 
simulations of the decaying shear waves problem, expressed with respect to the 
number of nodes $N_x$. 
A second second order convergence is recovered for the CSLB method, 
while for the FDLB method, the convergence is of fifth order.}
\label{error_analysis_sw}
\end{center}

\end{figure}

We fixed the value of the kinematic viscosity in FD units at $\nu_{0;\text{FD}}= \tau_{\text{FD}} = 5 \times 10^{-4}$, and the simulations were performed for various values of $N_x=N_{LU}$.
For a given value of $\nu_{\text{FD}}$ and number of lattice nodes $N_x$, we used Eq.~\eqref{eq:nu-transfer} 
to obtain the corresponding value of $\tau_{\text{LU}}$, in order to simulate the exact same system with the CS and FD models.
For the FD model we used a time step of $(\delta t)_{\text{FD}}=5\times10^{-4}$ 
and lattice spacing $(\delta s)_{\text{FD}}=1/N_x$. In the CS model, the time step 
$(\delta t)_{\text{LU}} = 1/\sqrt{3}$ corresponds to $(\delta t)_{\text{FD}} = 1 / N_x \sqrt{3} \simeq 
5 \times 10^{-4} \times (1155 / N_x)$, which for $20 \le N_x \le 60$ is around $20$ to $60$ times larger 
than the time step employed in the FDLB implementation.

In order to perform a quantitative analysis, a numerical fit of Eq.~\eqref{eq:shear_sol} was performed, 
which allows the parameter $\nu_{\rm app}$ to be extracted.
The measured values of $\nu_{app}$  are reported in Table \ref{tabel_shear} with the corresponding relative error. 
The latter is plotted in Fig. \ref{error_analysis_sw}, showing a second order convergence for CS and a fifth order one for FD
and confirming the expected numerical accuracy of the used models. 
It is worth noting that at $N_x= 20$, the relative error when the CSLB method 
is employed is roughly $3.5$ times smaller than the one corresponding to the FDLB method.
The relative error of the FDLB results becomes smaller than that corresponding to the 
CSLB method when $N_x \gtrsim 30$.

\subsection{Taylor-Green vortices}\label{sec:res:tg}
\begin{figure}
\centering
\includegraphics[width=0.7\textwidth]{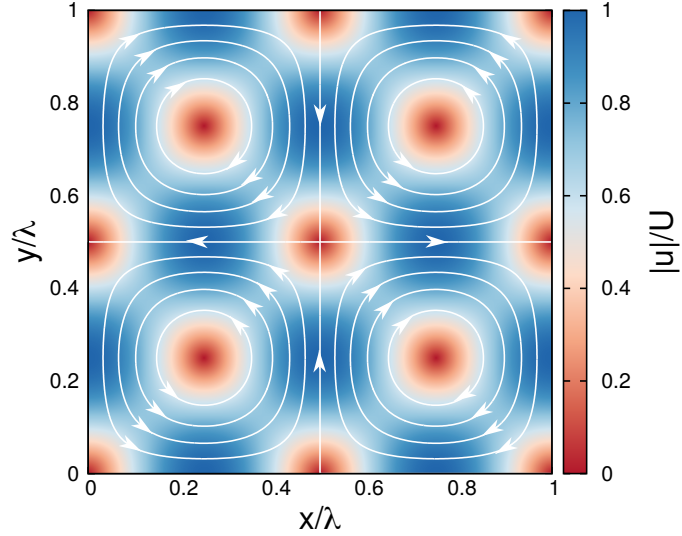}\\
\caption{Initial structure of a Taylor-Green vortex flow. Contour plot of the velocity field module is shown, with superimposed velocity stream lines.}
\label{taylor_green_conf}
\end{figure}

A parallel check for the kinematic viscosity can be performed by analyzing 
the damping of 2D Taylor-Green vortices. The system is initialized as follows:
\begin{eqnarray}
 u_x & = & U  \sin(kx) \cos(ky),\nonumber\\
 u_y & = & -U  \cos(kx) \sin(ky),
 \label{eq:tg}
\end{eqnarray}
where the amplitude is $U=0.01$ and the wave vector is $\bm{k} = (k, k)$, with 
$k = 2\pi / \lambda = 2\pi$. 

Similarly to the shear wave case, if we assume that for $t>0$, Eq. \eqref{eq:tg} holds with the 
amplitude $U$ replaced by $\widetilde{u}(t)$, then 
\begin{equation}
\widetilde{u}(t) = U e^{-2 k^2 \nu_0 t}.\label{eq:taylor}
\end{equation}
Fig. \ref{taylor_green_conf} shows the initial structure of a Taylor-Green vortex flow. The flow maintains the same structure
while decaying exponentially.

\begin{figure}
\begin{center}
\includegraphics[width=0.7\linewidth]{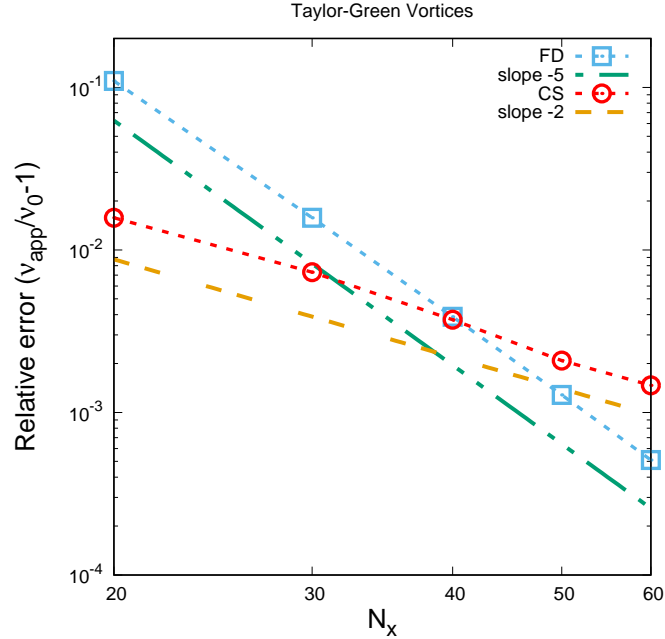}
\caption{Relative error of the measured kinematic viscosity $\nu_{app}$ expressed 
with respect to the expected analytic value $\nu_0$, extracted from the numerical 
simulations of the decaying Taylor-Green vortices, expressed with respect to the 
number of nodes $N_x$. 
A second order convergence is recovered for the CSLB method, 
while for the FDLB method, the convergence is of fifth order.}
\label{error_analysis_tg}
\end{center}
\end{figure}

We fixed again the value of the kinematic viscosity in FD units at $\nu_{0;FD}=5 \times 10^{-4}$, 
and the simulations were performed on square lattices having various number of nodes $N_x=N_y=N_{LU}=N$.
The measured values of $\nu_{app}$, obtained by numerically fitting the simulation results with Eq.~\eqref{eq:taylor},
are reported in Table~\ref{tabel_tg} alongside the corresponding relative error.
The latter is plotted in Fig.~\ref{error_analysis_tg}, showing again a 
second order convergence for CS and a fifth order one for FD with respect to the number of nodes.
At $N = 20$, the relative error obtained using the CSLB model is about $7$ times smaller 
than the one corresponding to the FDLB results. The relative error of the FDLB results 
becomes smaller than the corresponding CSLB error when $N_x \gtrsim 40$.

\begin{table}[ht!]
\tbl{Apparent kinematic viscosity $\nu_{app}$, expressed using the FD adimensionalization, measured as a numerical fit of Eq. (\ref{eq:taylor})
in the context of the damping of the Taylor-Green vortices. }
{\begin{tabular}{@{}|ll|ll|ll|@{}}
\hline
\multicolumn{2}{|l|}{\cellcolor[HTML]{656565}\ }   &
\multicolumn{2}{c|}{\cellcolor[HTML]{656565}{\color[HTML]{FFFFFF}CSLB}}   &
\multicolumn{2}{c|}{\cellcolor[HTML]{656565}{\color[HTML]{FFFFFF}FDLB}}   \\\hline
$\tau_{\text{LU}}$ & $N_x$ & $\nu_{app}$   & Rel. err. & $\nu_{app}$ & Rel. err. \\ 
$0.2986$    & $20$  & $0.0005078$ & $0.0157$  & $0.0005548$ & $0.1096$    \\
$0.3036$    & $30$  & $0.0005036$ & $0.0072$  & $0.0005079$ & $0.0158$    \\
$0.3086$    & $40$  & $0.0005018$ & $0.0037$  & $0.0005019$ & $0.0039$    \\
$0.3136$    & $50$  & $0.0005010$ & $0.0020$  & $0.0005006$ & $0.0013$    \\
$0.3186$    & $60$  & $0.0005007$ & $0.0014$  & $0.0005002$ & $0.0005$    \\ 
\hline
\end{tabular}
\label{tabel_tg}}
\end{table}

\section{Conclusions}\label{sec:conc}
We presented a comparison between lattice Boltzmann models implemented using 
the collision-streaming (CSLB) and finite-difference (FDLB) approaches.
By matching the physical parameters such as the kinematic viscosity and system size,
we showed how the results obtained using the two implementations can be related with each other,
despite the different mathematical and computational complexity of these two methods. 

We considered three different problems, namely the propagation of shock waves, 
the damping of shear waves and the damping of the Taylor-Green vortices.
A good agreement between the two models was observed when the simulation 
parameters were chosen to correspond to the same physical quantities.

By providing the tools necessary to control the relevant physical quantities
within the FD and CS approaches, our study confirms that is possible to simulate
the same physical system using these two approaches, thus paving the way to 
address in the future fluid systems for wider ranges of parameters.
This will be useful, e.g., in future simulations of nonideal fluids, 
since using both the CS and FD approaches can allow wider ranges of the 
parameter space to be explored.

\section{Acknowledgments}
V. E.  Ambru\c{s} and S. Busuioc  acknowledge funding from the Romanian Ministry of Research and Innovation, CCCDI-UEFISCDI, project number PN-III-P1-1.2-PCCDI-2017-0371/VMS, within PNCDI III.

\bibliographystyle{unsrt}
\bibliography{manuscript_Negro}

\end{document}